# Superconductivity and Crystal Structure of the Palladium-Iron-Arsenides $Ca_{10}(Fe_{1-x}Pd_xAs)_{10}Pd_3As_8$

C. Hieke[a], J. Lippmann[a], T. Stürzer[a], G. Friederichs[a], F. Nitsche[a],

F. Winter[b], R. Pöttgen[b] and D. Johrendt[a]*

[a]*Department Chemie der Ludwig-Maximilians-Universität München, Butenandtstr. 5-13 (Haus D), 81377 München, Germany;* [b]*Institut für Anorganische und Analytische Chemie, Universität Münster, Corrensstr. 30, 48149 Münster, Germany*

*Corresponding author. Email: dirk.johrendt@cup.uni-muenchen.de

The palladium-iron-arsenides $Ca_{10}(Fe_{1-x}Pd_xAs)_{10}(Pd_3As_8)$ were synthesized by solid state methods and characterized by X-ray powder and single crystal diffraction. The triclinic crystal structure (space group $P\bar{1}$) is isotypic to the homologue platinum 1038 type superconductors with alternating $FeAs_{4/4}$- and $Pd_3As_8$-layers, each separated by layers of calcium atoms. Iron is tetrahedral and palladium is planar coordinated by four arsenic atoms. $As_2$-dimers ($d_{As-As}$ ≈ 250 pm) are present in the $Pd_3As_8$-layer. Even though each layer itself has a fourfold rotational symmetry, the shifted layer stacking causes the triclinic space group. Resistivity measurements of La-doped samples show the onset of superconductivity at 17 K and zero resistivity below 10 K. The magnetic shielding fraction is about 20 % at 3.5 K. $^{57}$Fe-Mössbauer spectra exhibit one absorption line and show no hint to magnetic ordering. The electronic structure is very similar to the known iron-arsenides with cylinder-like Fermi surfaces and partial nesting between hole- and electron-like sheets. Our results show that superconductivity in the palladium-iron-compounds is present but complicated by too high substitution of iron by palladium in the active FeAs-layers. Since the electronic preconditions are satisfied, we expect higher critical temperatures in Pd1038-compounds with lower or even without Pd-doping in the FeAs-layer.

**Introduction**

Iron-arsenide and iron-selenide materials represent a new class of high-$T_c$ superconductors beyond the copper-oxides [1, 2]. Both materials have layer-like crystal structures, where the active layers are $CuO_2$ or $FeX$ ($X$ = As, Se), respectively. Superconductivity emerges in these active layers in the proximity of antiferromagnetic ordering that becomes suppressed by doping. It is widely accepted that the pairing in copper-oxides and iron-arsenides is unconventional, and many recent results indicate that magnetic fluctuations play a fundamental role [3].

Considering the fact that superconductivity develops in $CuO_2$- or $FeX$-layers, the large variations of the critical temperatures in compounds with different crystal structures yet the same active layers are as remarkable as poorly understood. It is obvious that the surrounding part of the crystal structure plays a crucial role beyond the function as charge reservoir. However, a more detailed understanding has yet to be achieved.

Until 2011 the family of iron-arsenide superconductors consisted of compounds with relatively simple and well known types of crystal structures, mostly variants of the PbFCl- and $ThCr_2Si_2$-types. Therein, the negatively charged FeAs-layers are separated and charge balanced by layers of alkaline- or alkaline-earth ions like in $BaFe_2As_2$ or NaFeAs, or by positively charged layers of rare-earth-oxide like in LaOFeAs. A certain expansion provided the compounds where the FeAs layers are separated by thicker perovskite-like oxide-blocks as in $Sr_2VO_3FeAs$ [4], which is isostructural to the copper sulphide $Sr_2GaO_3CuS$. All these separating layers are itself insulating or semiconducting. By combination with the metallic FeAs-layer the Fermi-surface remains unaffected, as soon as the Fermi level remains in the gap of electronic states of the separating layers. This is always the case if the gap is large enough (as it usually is in oxides), and naturally if the separating layers consist only of very electropositive alkaline- or alkaline-earth atoms.

Recently new iron-based superconductors were found where the FeAs-layers are separated by calcium atoms and semiconducting negatively charged platinum-arsenide-layers [5-7]. The compounds $Ca_{10}(FeAs)_{10}Pt_3As_8$ (referred to as 1038) and $Ca_{10}(FeAs)_{10}Pt_4As_8$ (referred to as 1048) exhibit superconductivity up to 38 K and have raised the chemical and structural complexity of the iron-arsenide family. Recently we have shown that triclinic $Ca_{10}(FeAs)_{10}Pt_3As_8$ is the non-superconducting and magnetically ordered parent compound of these materials [8]. Superconductivity can be induced by Pt-doping of the FeAs-layers ($Ca_{10}(Fe_{1-x}Pt_xAs)_{10}Pt_3As_8$, $T_c \leq 15$ K) or by electron-doping either via La-substitution ($(Ca_{1-x}La_x)_{10}(FeAs)_{10}Pt_3As_8$, $T_c \approx 35$ K) or by charge transfer from the platinum-arsenide-layer ($Ca_{10}(FeAs)_{10}Pt_4As_8$, $T_c \approx 38$ K) [9]. Already from the relatively high critical temperatures one may assume that the Fermi-surface (FS) originates from the FeAs-layers and is scarcely affected by states from the platinum-arsenide layers. This is supported by band structure calculations [5], photoemission experiments [10] and specific heat data [11]. In other words, the Fermi-energy of the metallic FeAs-layer lies just in the small gap of the semiconducting $Pt_3As_8$-layer. This is inherently remarkable, and moreover on precondition for the high critical temperatures which are believed to be tied to a special FS topology generated by the FeAs layer alone.

In face of this special situation, one might assume that the $Ca_{10}(FeAs)_{10}Pt_3As_8$-type superconductors (Pt1038) are rather unique and possibly intolerant against substitution of the $Pt_zAs_8$-layers. In this letter we show that this is not the case, and report the synthesis, crystal structure, superconducting properties and electronic structure of the palladium-iron-arsenides $Ca_{10}(Fe_{1-x}Pd_xAs)_{10}Pd_3As_8$ referred to as Pd1038 compounds.

**Results**

*Synthesis and crystal structure*

Polycrystalline samples of the palladium-iron-arsenides were synthesized by solid state methods. Stoichiometric mixtures of pure elements (>99.5%) were heated at 1000 °C in alumina crucibles sealed in silica tubes under purified argon. X-ray powder patters were similar to the Pt1038 compounds suggesting an isotypic crystal structure. In order to confirm this, a small single crystal (12×5×1 μm$^3$) was selected for X-ray single crystal structure determination. Reflection data processing and structure refinement turned out difficult due to diffuse scattering along the stacking direction, and additionally to partial merohedral twinning. Structure refinement was successful in space group $P\bar{1}$ by using four twin domains with the Jana program package [12]. The tetragonal layer symmetry of the FeAs and $Ca_{10}Pd_3As_8$ substructures is reduced to inversion in the 3D arrangement due to mismatched stacking. The square basal plane of the triclinic cell ($a = b$; $\gamma = 90°$) still reflects the tetragonal motifs of the layers. As a consequence, partial merohedral twinning occurs where reflections of all domains with $2h+k = 5n$ coincide. Results of the crystal structure refinement are compiled in Table 1 together with selected interatomic distances and angles within the $Fe_{1-x}Pd_xAs$- and $Pd_3As_8$-layers.

The crystal structure of $Ca_{10}(Fe_{0.84}Pd_{0.16}As)_{10}Pd_{2.8}As_8$ is depicted in Fig. 1. The bond lengths within the tetrahedral $Fe_{1-x}Pd_xAs$-layers (red tetrahedra) match with typical values around 240 pm known from other iron-arsenide superconductors [13, 14]. However, we observe slightly longer distances up to 247 pm if the palladium-doping level at the iron-sites increases. The bond angles within the tetrahedra are all close to the ideal value of 109.4° which is believed favorable for high critical temperatures [15]. The $Pd_3As_8$-layer consists of corner-sharing $PdAs_{4/4}$-squares where the arsenic atoms form

As$_2$-dimes with As–As bond lengths of 248 and 250 pm, which is slightly longer than twice the covalent radius of arsenic that is 242 pm [16].

Table 1: Crystal data and structure refinement for Pd1038, space group $P\bar{1}$, $Z = 2$.

| | |
|---|---|
| Empirical formula | Ca$_{10}$(Fe$_{0.84}$Pd$_{0.16}$As)$_{10}$Pd$_{2.8}$As$_8$ |
| Molar mass | 2685.5 g mol$^{-1}$ |
| Unit cell dimensions (single crystal) | $a = 880.0(1)$ pm |
| | $b = 880.0(1)$ pm |
| | $c = 1060.3(2)$ pm |
| | $\alpha = 85.271(2)°$ |
| | $\beta = 75.561(2)°$ |
| | $\gamma = 90.003(6)°$ |
| | $V = 0.7922(1)$ nm$^3$ |
| Calculated density | 5.63 g cm$^{-3}$ |
| Crystal size | $12 \times 5 \times 1$ µm$^3$ |
| Wave length | 71.073 pm (MoK$_\alpha$) |
| Transmission ratio (max / min) | 0.1341 / 0.0229 |
| Absorption coefficient | 26.45 mm$^{-1}$ |
| $\Theta$ range | 1.99 - 30.31° |
| Range in $hkl$ | ± 12, ± 12, ± 15 |
| Total number reflections | 121430 |
| Independent reflections / $R_{int}$ | 27887 / 0.1369 |
| Reflections with $I \geq 3\sigma(I)$ / $R_\sigma$ | 7090 / 0.1239 |
| Data / parameters | 7090 / 147 |
| Goodness-of-fit on $F$ | 4.29 |
| $R1$ / $wR2$ for $I \geq 3\sigma(I)$ | 0.0883 / 0.986 |
| $R1$ / $wR2$ for all data | 0.0883 / 0.986 |
| Largest diff. peak and hole | 8.82 / −3.43 eÅ$^{-3}$ |

Interatomic distances and angles:

Fe$_{1-x}$Pd$_x$As-layer  
$d_{(Fe/Pd–As)} = 237.3(4)$ - $247.5(4)$ pm

∡$_{(As–Fe/Pd–As)} = 107.9(4)$ - $110.1(4)°$

Pd$_3$As$_8$-layer  
$d_{(Pd–As)} = 235.0(2)$ - $258.6(2)$ pm  
$d_{(As–As)} = 248.5(2), 250.6(2)$ pm

∡$_{(As–Pd–As)} = 85.2(5)$ - $94.8(5)°$

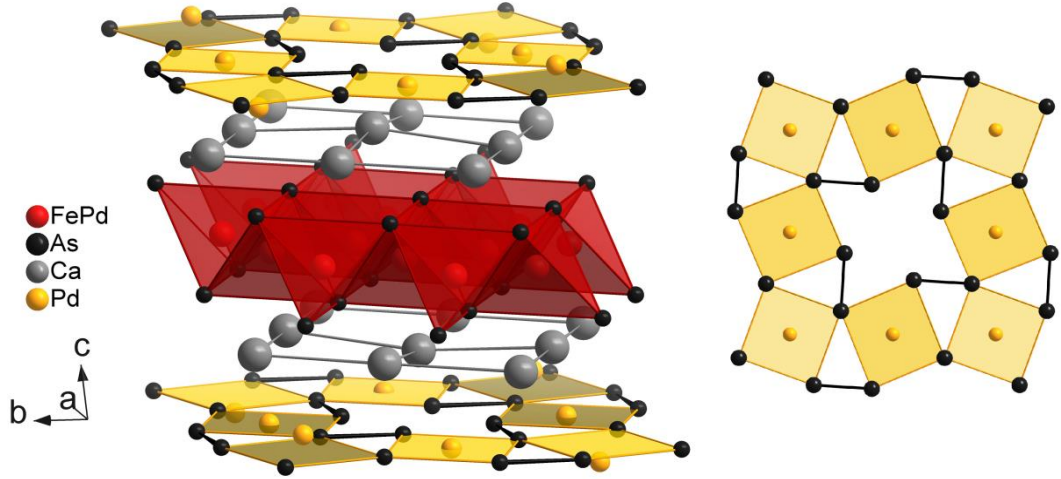

Figure 1: Crystal structure of $Ca_{10}(Fe_{0.84}Pd_{0.16}As)_{10}Pd_{2.8}As_8$. Left: stacking of the active $Fe_{1-x}Pd_xAs$- with the $Pd_3As_8$-layers, each separated by calcium atoms. Right: View of the $Pd_3As_8$-layer emphasizing the square coordination of palladium and the $As_2$-dimers.

By assuming an idealized composition $Ca_{10}(FeAs)_{10}Pd_3As_8$ and typical oxidation states, we arrive at the charge balanced formula $Ca_{10}^{2+}(Fe^{2+}As^{3-})_{10}(Pd_3^{2+}As_8^{2-})$. The charge of the arsenic atoms in the $Pd_3As_8$-layer is 2− due to the homonuclear bonds of the $As_2$-dimers, thus the parent Pd1038-phase ranks among the Zintl-phases. Note that the iron-arsenide layer in $Ca_{10}(FeAs)_{10}Pd_3As_8$ carries one negative charge exactly as the known parent compounds $LaOFeAs$, $BaFe_2As_2$ or $NaFeAs$. So far we were not able to synthesize the pure parent compound without Pd-substitution at the iron site. Furthermore the single crystal structure refinement indicated a small Pd-deficiency in the $Pd_3As_8$-layer according to $Pd_{2.8}As_8$. Initially the homogeneity of polycrystalline samples was rather poor, but could be improved by additional La-doping at the Ca-position. This effect was known from the Pt1038 compounds and is probably caused by the increased lattice energy if $La^{3+}$-ions replace the lower charged $Ca^{2+}$-ions at certain positions [9]. Figure 2 shows the X-ray powder pattern and Rietveld refinement (program Topas [17]) of a La-doped sample with composition $(Ca_{0.92}La_{0.08})_{10}(Fe_{0.86}Pd_{0.14}As)_{10}Pd_{2.8}As_8$. This profile fit bases on the structural data obtained from the single-crystal experiment, and attests a polycrystalline sample of the La-doped Pd1038 phase with a minor impurity of FeAs ($\approx$ 10 %).

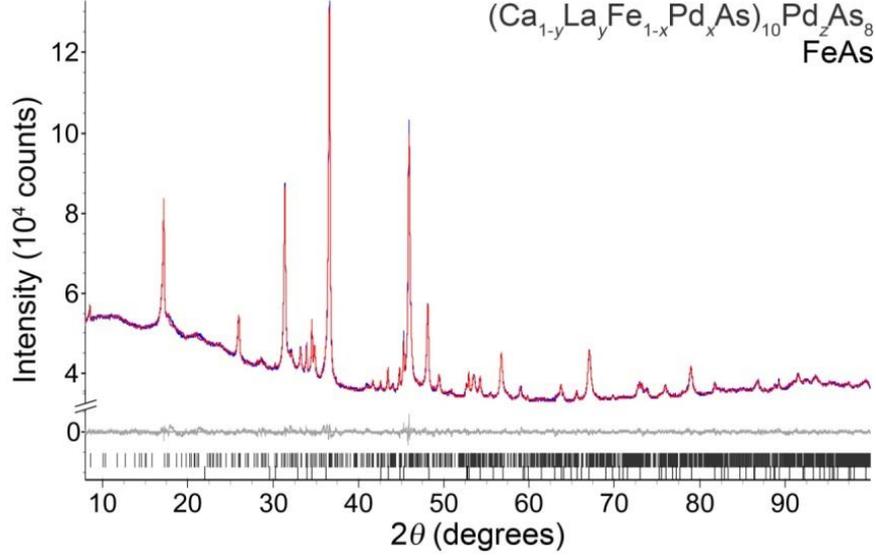

Figure 2: X-ray diffraction pattern (blue) and Rietveld fit (red) of $(Ca_{0.92}La_{0.08})_{10}(Fe_{0.86}Pd_{0.14}As)_{10}Pd_{2.8}As_8$ ($P\bar{1}$, $a$ = 880.57(3) pm, $b$ = 881.34(2) pm, $c$ = 1063.9(3) pm, $\alpha$ = 85.239(1)°, $\beta$ = 75.653(2)°, $\gamma$ = 89.879(2)°, $R_{wp}$ = 0.0086).

*Resistivity and magnetic susceptibility*

Figure 3 shows the dc-resistivity of the La-doped Pd1038 sample. The cold pressed pellet reveals the onset of a superconducting transition at 17 K while zero resistivity is achieved below 10 K. This is in agreement with the ac-susceptibility measurement shown in the inset. The shielding fraction is about 20% at 3.5 K and not fully developed at this temperature. Thus we expect larger shielding fractions at lower temperatures according to bulk superconductivity. The rather broad superconducting transition together with the weak shielding is very probable a consequence of inhomogeneous distribution of the different dopants. This is not surprising because a homogeneous distribution of Ca/La- and Fe/Pd-substitution together with Pd-deficiency in the $Pd_3As_8$-layer is certainly hard to achieve. Nevertheless our results prove that superconductivity emerges in the Pd1038 compound even though the critical temperature is still lower when compared with the Pt1038 materials. We actually would not expect superconductivity if the composition is $(Ca_{0.92}La_{0.08})_{10}(Fe_{0.86}Pd_{0.14}As)_{10}Pd_{2.8}As_8$ throughout the sample. From the phase diagrams of $Ba(Fe_{1-x}Pd_x)_2As_2$ [18] and the Pt1038 compounds $Ca_{10}(Fe_{1-x}Pt_xAs)_{10}Pt_3As_8$ [9, 19] we know that superconductivity is induced by Pd- or Pt-doping, but only at low doping levels up to about 8 %. On the other hand, we have shown that superconductivity in the Pt1038-phase can be induced by La-doping if Pt-doping at the Fe-site is absent [9]. Thus our results suggest that the sample contains fractions where the Pd-substitution is low enough to allow superconductivity, which also explains the reduced shielding fraction.

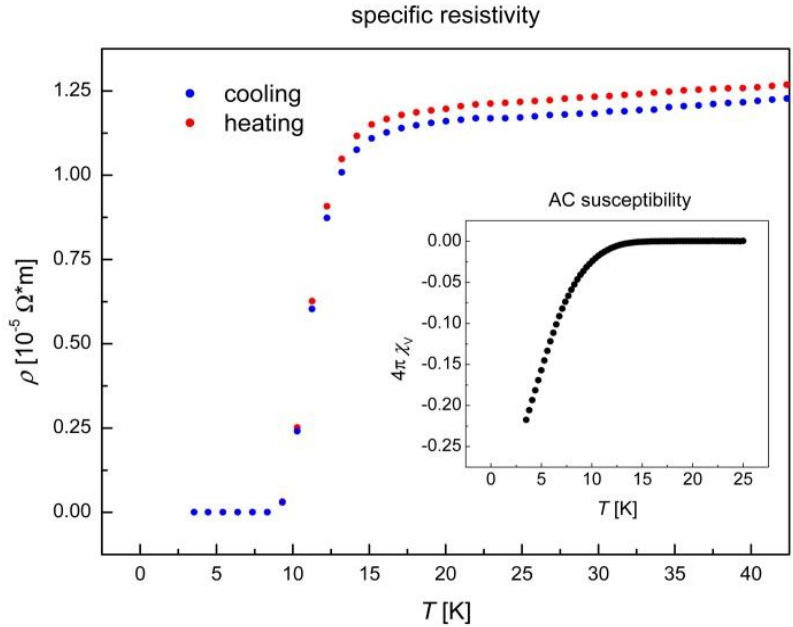

Figure 3: *dc*-Resistivity of polycrystalline $(Ca_{0.92}La_{0.08})_{10}(Fe_{0.86}Pd_{0.14}As)_{10}Pd_{2.8}As_8$. Insert: ac-susceptibility measured at 3 Oe and 1333 Hz.

## $^{57}$Fe Mössbauer spectroscopy

The $^{57}$Fe Mössbauer spectra for the $(Ca_{0.87}La_{0.13})_{10}(Fe_{0.86}Pd_{0.14}As)_{10}Pd_{2.8}As_8$ sample collected at various temperatures are shown in Figure 4 together with transmission integral fits [20]. The corresponding fitting parameters are listed in Table 2. The three spectra are well reproduced with single signals, however, with slightly enhanced line widths, especially towards lower temperature. The iron atoms are distributed over five crystallographically independent sites. Thus we observe a distribution of five subspectra and the experimental one is the superposition. Since the parameters of the subspectra are very close, we only observe the envelope curve. The line width enhancement in the 5 K spectrum can be explained by a distribution of different distortions of the $Fe_{1-x}Pd_xAs_4$ tetrahedra. This is also expressed by the slightly enhanced quadrupole splitting parameter (Table 2). The isomer shift (due to the superposition we get an average value) increases from 0.31 mm/s at 293 to 0.45 mm/s at 5 K, a consequence of a second order Doppler shift (SODS). The absolute $\delta$ values compare well with other iron arsenides, e. g. the solid solution $Ba_{1-x}K_xFe_2As_2$ [21, 22] or $Sr_3Sc_2O_5Fe_2As_2$ [23]. The 78 K and 5 K spectra give no hint for a hyperfine field contribution thus no magnetic ordering is present.

Table 2: Fitting parameters of $^{57}$Fe Mössbauer spectra of $(Ca_{0.87}La_{0.13})_{10}(Fe_{0.86}Pd_{0.14}As)_{10}Pd_{2.8}As_8$ at different temperatures. $\Gamma$: experimental line width, $\delta$: isomer shift; $\Delta E_Q$: electric quadrupole splitting parameter (for details see text).

| $T$ (K) | $\delta$ (mm s$^{-1}$) | $\Delta E_Q$ (mm s$^{-1}$) | $\Gamma$ (mms$^{-1}$) |
|---|---|---|---|
| 293 | 0.31(1) | 0.21(3) | 0.45(4) |
| 78 | 0.43(1) | 0.34(6) | 0.91(6) |
| 5 | 0.45(2) | 0.54(6) | 1.23(8) |

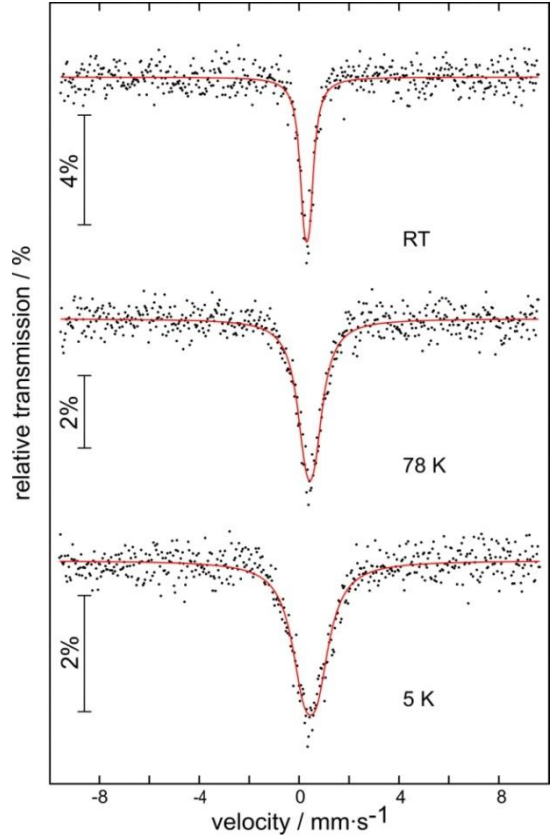

Figure 4: Experimental and simulated $^{57}$Fe Mössbauer spectra ($^{57}$Co/Rh source) of $(Ca_{0.87}La_{0.13})_{10}(Fe_{0.86}Pd_{0.14}As)_{10}Pd_{2.8}As_8$ at various temperatures.

*Electronic structure calculation*

The electronic structure of $Ca_{10}(FeAs)_{10}Pd_3As_8$ was calculated by the full-potential LAPW-lo method [24] using the WIEN2k package [25, 26]. Figure 5a shows the total electronic density of states (DOS) together with partial DOS projections (pDOS) of the FeAs- and $Pd_3As_8$-layers. The FeAs pDOS (red curve) is very similar to other iron-arsenide materials, which suggests almost independent electronic systems, i.e. very weak coupling between iron-arsenide and palladium-arsenide layers in $Ca_{10}(FeAs)_{10}Pd_3As_8$. The states at the Fermi-energy ($E_F$) are dominated by FeAs, even though a certain contribution of orbitals from the $Pd_3As_8$-layer (blues line) is discerni-ble. From the pDOS values at $E_F$ we estimate 86% FeAs- and 14% $Pd_3As_8$-states, respectively, while calcium-orbitals are negligible at $E_F$.

The calculated Fermi surface (FS) topology of triclinic $Ca_{10}(FeAs)_{10}Pd_3As_8$ is depicted in the Figures 5b-f. For clarity we show the different sheets (Figures b-e) and the complete FS (Figure 5f) separately. Cylinders along $c^*$ around the Γ-point (Figure 5b+c) and the zone corners (Figure 5c) show the two-dimensionality of the electronic structure, and strongly resembles the FS topology of other iron-arsenide superconductors like $CaFe_2As_2$ shown in Figure 5g for comparison. Nesting between hole-like and electron-like FS sheets with a momentum vector $q = $ (½ ½ 0) is intensively discussed as being essential for high-$T_c$ in iron-arsenides [27-30]. Spin fluctuations with this momentum vector were found experimentally [31, 32], and they are now thought to play a key role for the pairing mechanism [3]. Indeed significant parts of the FS in Figure 5f nearly coincide when shifted by (½ ½ 0), and qualifies $Ca_{10}(FeAs)_{10}Pd_3As_8$ as typical iron-arsenide material from the view of the electronic structure. However, we also observe some deviations of the FS as seen in Figure 5e, where the cylinders become a torus-like shape that is not present in the other FeAs compounds. However, these perturbations are much weaker than those calculated for $Sr_2VO_3FeAs$ by assuming non-magnetic vanadium [33, 34].

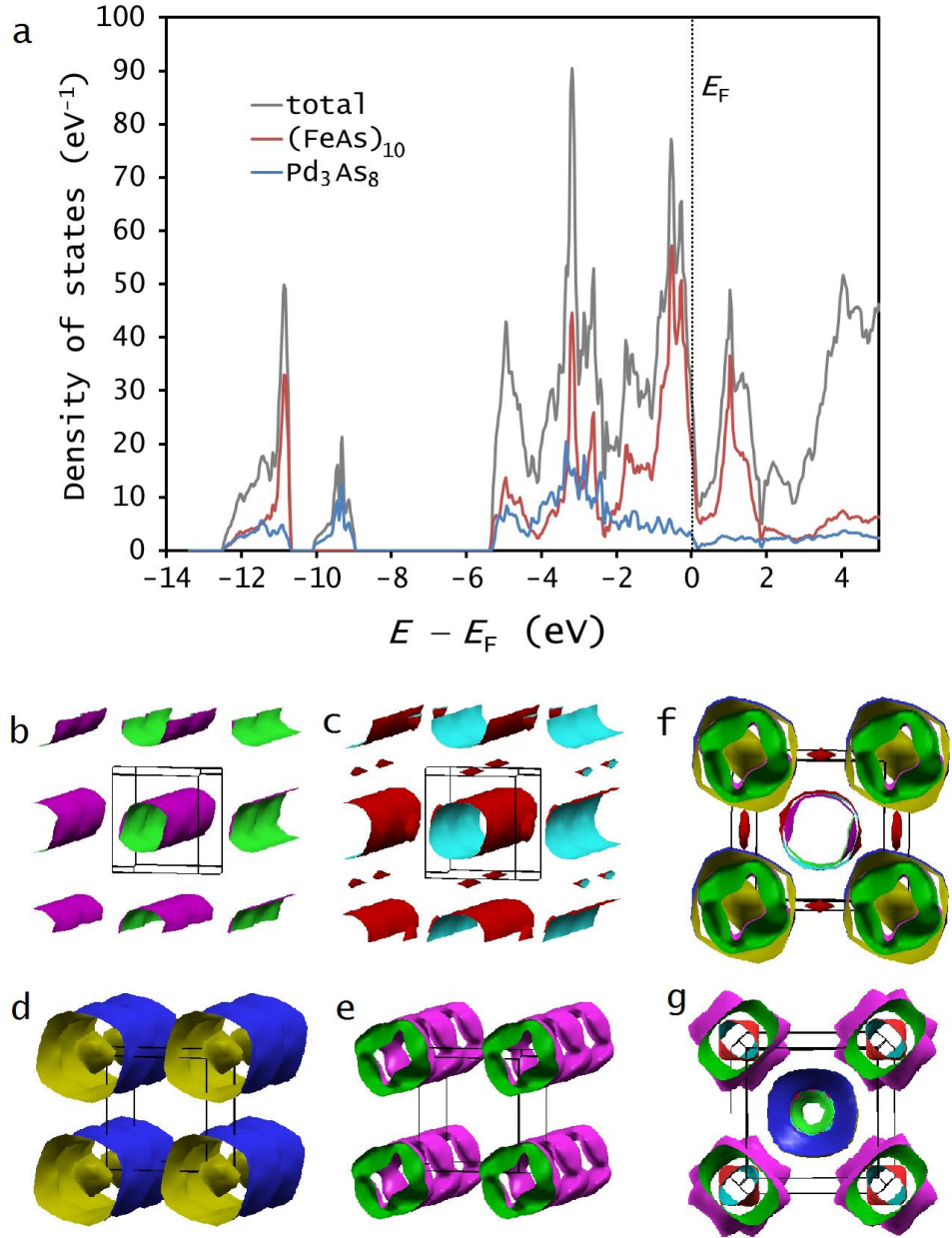

Figure 5. a: Electronic density of states of (DOS) $Ca_{10}(FeAs)_{10}Pd_3As_8$; Total DOS (grey), contributions of the FeAs- (red) and $Pd_3As_8$-layers (blue). b-e: Fermi surface sheets, f: sheets b-e merged, g: Fermi surface of $CaFe_2As_2$.

**Conclusion**

We have shown that 1038-type superconductors exist with palladium-arsenide layers. The single crystal X-ray structure determination confirmed the triclinic crystal structure of $Ca_{10}(Fe_{1-x}Pd_xAs)_{10}Pd_3As_8$ isotypic to Pt1038, even though the data accuracy is slightly lowered due to partial merohedral twinning and diffuse scattering. Polycrystalline samples of $(Ca_{0.92}La_{0.08})_{10}(Fe_{0.86}Pd_{0.14}As)_{10}Pd_{2.8}As_8$ reveal the onset of a superconducting transition at 17 K and still incomplete magnetic shielding of about 20%. The relative low $T_c$ when compared with the Pt-compounds and the relatively weak shielding are ascribed to sample issues, where Pd-overdoping at the Fe-site is detrimental to superconductivity. We suggest that the inhomogeneous distribution of the Pd-atoms in the Fe-layer produces certain fractions that have Pd-concentrations low enough to allow superconductivity. $^{57}$Fe Mössbauer data agree well with other FeAs superconductors and show no hints of magnetic ordering. Electronic structure calculations of the idealized parent compound $Ca_{10}(FeAs)_{10}Pd_3As_8$ reveal weak coupling between the FeAs- and $Pd_3As_8$-layers, and a FS topology very similar to the known FeAs superconductors. Typical FS features like cylinders at Γ and X as well as partial FS nesting qualifies the new compounds as typical iron-arsenide superconductors. From this we conclude that the Pd1038-compounds will exhibit higher critical temperatures in samples with lower or ideally without Pd-doping in the FeAs-layer.

**Acknowledgements**

This work was financially supported by the German Research Foundation (DFG) within the priority program SPP1458 under grant JO257/6-2.

**References**

[1]     G.R. Stewart, Rev. Mod. Phys. 83 (2011) p.1589.

[2]     D. Johrendt and R. Pöttgen, Angew. Chem. Int. Ed.  47 (2008) p.4782.

[3]     D.J. Scalapino, Rev. Mod. Phys. 84 (2012) p.1383.

[4]     X. Zhu, F. Han, G. Mu, P. Cheng, B. Shen, B. Zeng and H.-H. Wen, Phys. Rev. B 79 (2009) p.220512.

[5]     C. Löhnert, T. Stürzer, M. Tegel, R. Frankovsky, G. Friederichs and D. Johrendt, Angew. Chem. Int. Ed. 50 (2011) p.9195.

[6]     N. Ni, J.M. Allred, B.C. Chan and R.J. Cava, Proc. Natl. Acad. Sci. U. S. A. 108 (2011) p.E1019.

[7]     S. Kakiya, K. Kudo, Y. Nishikubo, K. Oku, E. Nishibori, H. Sawa, T. Yamamoto, T. Nozaka and M. Nohara, J. Phys. Soc. Jpn. 80 (2011) p. 093704

[8]     T. Stürzer, G.M. Friederichs, H. Luetkens, A. Amato, H.H. Klauss and D. Johrendt, J. Phys.: Condens. Matter 25 (2013) p.122203.

[9]     T. Stürzer, G. Derondeau and D. Johrendt, Phys. Rev. B 86 (2012) p.060516(R).

[10]    M. Neupane, C. Liu, S.-Y. Xu, Y.J. Wang, N. Ni, J.M. Allred, L.A. Wray, H. Lin, R.S. Markiewicz, A. Bansil, R.J. Cava and M.Z. Hasan, arxiv:1110.4687 (2011).

[11]    J.S. Kim, T. Stürzer, D. Johrendt and G.R. Stewart, J. Phys.: Condens. Matter 25 (2013) p.135701.

[12]    V. Petricek, M. Dusek and L. Palatinus *Jana2006. Structure Determination Software Programs*, Institute of Physics, 2009.

[13]    M. Rotter, M. Tegel, I. Schellenberg, W. Hermes, R. Pöttgen and D. Johrendt, Phys. Rev. B 78 (2008) p.020503(R).

[14]    M. Tegel, S. Johansson, V. Weiß, I. Schellenberg, W. Hermes, R. Pöttgen and D. Johrendt, Europhys. Lett. 84 (2008) p.67007.

[15]    C.H. Lee, A. Iyo, H. Eisaki, H. Kito, M.T. Fernandez-Diaz, T. Ito, K. Kihou, H. Matsuhata, M. Braden and K. Yamada, J. Phys. Soc. Jpn. 77 (2008) p.083704.

[16]    L. Pauling, *The Nature of the Chemical Bond and the Structure of Molecules and Crystals: An Introduction to Modern Structural Chemistry*, Cornell University Press, Ithaca, NY, 1960.

[17]    A. Coelho, *TOPAS-Academic, Version 4.1, Coelho Software*, Brisbane, 2007.

[18] N. Ni, A. Thaler, A. Kracher, J.Q. Yan, S.L. Bud'ko and P.C. Canfield, Phys. Rev. B 80 (2009) p.024511.

[19] K. Cho, M.A. Tanatar, H. Kim, W.E. Straszheim, N. Ni, R.J. Cava and R. Prozorov, Phys. Rev. B 85 (2012) p.020504.

[20] R.A. Brand, *Normos Mössbauer fitting Program*, Universität Dortmund, 2002.

[21] M. Rotter, M. Tegel, I. Schellenberg, F.M. Schappacher, R. Pöttgen, J. Deisenhofer, A. Gunther, F. Schrettle, A. Loidl and D. Johrendt, New J. Phys. 11 (2009) p.025014.

[22] D. Johrendt and R. Pöttgen, Physica C 469 (2009) p.332.

[23] M. Tegel, I. Schellenberg, F. Hummel, R. Pöttgen and D. Johrendt, Z. Naturforsch. B 64 (2009) p.815.

[24] G.K.H. Madsen, P. Blaha, K. Schwarz, E. Sjostedt and L. Nordstrom, Phys. Rev. B 64 (2001) p.195134/1.

[25] P. Blaha, K. Schwarz, G.K.H. Madsen, D. Kvasnicka and J. Luitz *WIEN2k – An Augmented Plane Wave + Local Orbitals Program for Calculating Crystal Properties*, Technische Universität Wien, 2001.

[26] K. Schwarz and P. Blaha, Comput. Mat. Sci. 28 (2003) p.259.

[27] I.I. Mazin, D.J. Singh, M.D. Johannes and M.H. Du, Phys. Rev. Lett. 101 (2008) p.057003.

[28] K. Terashima, Y. Sekiba, J.H. Bowen, K. Nakayama, T. Kawahara, T. Sato, P. Richard, Y.-M. Xu, L.J. Li, G.H. Cao, Z.-A. Xu, H. Ding and T. Takahashi, Proc. Natl. Acad. Sci. U. S. A. 106 (2009) p.7330.

[29] I.I. Mazin, Nature 464 (2010) p.183.

[30] P. Richard, T. Sato, K. Nakayama, T. Takahashi and H. Ding, Rep. Prog. Phys. 74 (2011) p.124512.

[31] A.D. Christianson, E.A. Goremychkin, R. Osborn, S. Rosenkranz, M.D. Lumsden, C.D. Malliakas, I.S. Todorov, H. Claus, D.Y. Chung, M.G. Kanatzidis, R.I. Bewley and T. Guidi, Nature 456 (2008) p.930.

[32] P. Dai, J. Hu and E. Dagotto, Nature Phys 8 (2012) p.709.

[33] K.W. Lee and W.E. Pickett, Europhys. Lett. 89 (2010) p.57008.

[34] M. Tegel, T. Schmid, T. Stürzer, M. Egawa, Y.X. Su, A. Senyshyn and D. Johrendt, Phys. Rev. B 82 (2010) p.140507.